\documentclass[reprint,nofootinbib, amsmath,amssymb, aps, floatfix, 
twoside,superscriptaddress]{revtex4-1}

\usepackage[utf8]{inputenc}
\usepackage[brazilian]{babel}
\usepackage[T1]{fontenc}

\usepackage{mathtools}
\usepackage{amsmath}
\usepackage{amsfonts}
\usepackage{amssymb}
\usepackage{graphicx}
\usepackage{color}
\usepackage[colorlinks=true,citecolor=blue,linkcolor=blue,urlcolor=blue]{hyperref}
\usepackage{times}
\usepackage{amstext}
\usepackage{latexsym}
\usepackage[running,mathlines]{lineno}
\usepackage{verbatim}
\usepackage{physics}
\usepackage{bm}
\usepackage{ulem}

\usepackage{listings}
\usepackage{xcolor}

\definecolor{codegreen}{rgb}{0,0.6,0}
\definecolor{codegray}{rgb}{0.5,0.5,0.5}
\definecolor{codepurple}{rgb}{0.58,0,0.82}
\definecolor{backcolour}{rgb}{0.95,0.95,0.92}

\lstdefinestyle{mystyle}{
    backgroundcolor=\color{backcolour},   
    commentstyle=\color{codegreen},
    keywordstyle=\color{magenta},
    numberstyle=\tiny\color{codegray},
    stringstyle=\color{codepurple},
    basicstyle=\ttfamily\footnotesize,
    breakatwhitespace=false,         
    breaklines=true,                 
    captionpos=b,                    
    keepspaces=true,                 
    numbers=left,                    
    numbersep=5pt,                  
    showspaces=false,                
    showstringspaces=false,
    showtabs=false,                  
    tabsize=2
}

\begin{document}
\raggedbottom

\title{\Large{Feixe parcialmente coerente: Teoria e simulação} \\ \small{(Partially coherent beam: Theory and simulation)}}

\author{Gubio G. de Lima}\email[Endereço de correspondência:]{ gubiofisika@gmail.com}
\affiliation{Universidade Federal de São Carlos, Departamento de Física, São Carlos, SP, Brasil}

\author{Sinara S. Dourado}\email[Endereço de correspondência:]{sinara.dourado@estudante.ufscar.br}
\affiliation{Universidade Federal de São Carlos, Departamento de Física, São Carlos, SP, Brasil}

\begin{abstract} 
    Desde do desenvolvimento do laser buscamos continuamente o desenvolvimento das técnicas e teoria para obter feixes com alto grau de coerência, pois as fontes de luz naturais fornecem luz incoerente. No entanto, há aplicações em que é vantajoso usar os feixes parcialmente coerentes (PC) de forma controlada, como na propagação em meios turbulentos. Para obter um feixe PC no laboratório ou em simulações, precisamos de teorias e métodos específicos. Neste artigo, descrevemos uma introdução à teoria dos feixes PC, como gera-los por meio de decomposições modais e um guia passo-a-passo para simulação e análise da geração dos feixes, inspirado em experimentos. Para ilustrar os métodos, apresentamos como exemplo os feixes do tipo \textit{Gaussian Schell-Model}.

\begin{description}
\item[Palavras-chave] Feixes parcialmente coerentes. Decomposição modal. Gaussian Schell-Model.
\end{description}

\bigbreak 

    Since the development of lasers, we have continuously sought to advance techniques and theory to obtain beams with a high degree of coherence, as natural light sources provide incoherent light. However, there are applications where it is advantageous to use partially coherent (PC) beams in a controlled manner, such as in propagation through turbulent media. To generate a PC beam in the laboratory or in simulations, specific theories and methods are required. In this article, we provide an introduction to PC beam theory, describing how to generate them through modal decompositions and a step-by-step guide for simulating and analyzing beam generation inspired by experiments. To illustrate the methods, we present Gaussian Schell-Model beams as an example.. 

\begin{description}
\item[Keywords] Partially coherent beams. Modal decomposition. Gaussian Schell-Model beams.
\end{description}
\end{abstract}

\maketitle


\section{Introdução}


 A física estuda a luz e seus fenômenos há décadas em várias perspectivas, mas a teoria de coerência e óptica estatística foi desenvolvida na metade do século XX \cite{introdphotonic}. A coerência pode ser considerada como consequência das correlações entre os componentes do campo elétrico flutuante em dois ou mais pontos, assim  feixe de luz com baixa coerência é denominado feixe parcialmente coerente.  As fontes de luz naturais geram campos com baixa coerência e por muitos anos houve um buscar em desenvolver a teoria e geração de fontes de luz coerentes, como as produzidas por lasers. Contudo, em algumas aplicações tecnológicas, é desejável trabalhar com feixes com baixa coerência, que podem ser gerados e manipulados de maneira controlada, para ser aplicadas em litografia \cite{litografia} , sensoriamento remoto \cite{sensoriamento_remoto} e comunicação óptica no espaço livre \cite{comunicação_optica1},\cite{comunicação_optica2}. Além de gerar com controle devemos ser obtê-los de forma rápida para transmitir informações e poder aplica-las em sistema de criptografia citação\cite{criptografia}. A chave para a manipulação dos feixes escalares, são suas propriedade de correlações espaciais  geralmente descrita pela intensidade mutual no domínio do espaço-tempo ou pela densidade espectral cruzada (CSD)  no domínio do espaço-frequência \cite{TRIBUTOWOFL,wolf2007introduction,mandel1995optical}. Permitindo o controle do perfil de intensidade no campo distante com intermédio de moduladores espaciais de luz \cite{Wang:20}. 
 
 O modelo mais básico e estudado de feixe parcialmente coerente é denominado  Gaussiano Schell (GSM) \cite{Tervonen_Friberg}, cuja intensidade e grau de coerência satisfazem as distribuições gaussianas. 
 Primeiramente introduzido na década de 1970 \cite{Wolf_Collett} e gerados pela primeira vez por De Santis et. al(1979) \cite{De_Santis} com disco giratório de vidro fosco (RGGD) e um filtro de amplitude gaussiana. Também Foi demonstrado que um feixe GSM tem vantagens sobre um feixe gaussiano totalmente coerente, em meios turbulento \cite{Gbur_wolf}.  O feixe GSM também pode ser gerado por uma superposição adequada de feixes de luz coerentes, mas não correlacionados entre si \cite{decomposição1,decomposição2,Wang:21}.

 Este trabalho visa apresentar uma introdução da teoria e conceitos por trás dos feixes parcialmente coerentes utilizando como exemplos as fonte do tipo GSM. No primeiro capitulo II iremos nos dedicar a falar sobre os conceitos matemáticos de graus de coerência e CSD. No capitulo III, iremos aborda os métodos de decomposição em modos que é utilizado para gerar os feixes e aplica-lo para os feixe GSM . Por fim, será apresentado um passo-a-passo de como simular os feixes, obter alguns parâmetros para comparar com a teoria. Este trabalho também visa apresentar uma alternativa para quem não tem acesso a ambientes experimentais com mesas óticas adequada , mais ainda sim querem investigar e estudar tópicos da área relacionado a fenômenos ópticos .


\section{Teoria: Feixes parcialmente coerentes }\label{sec2}


 \subsection{Grau de coerência e visibilidade }
 

Em livros-texto de óptica é comum abordar o conceito de campo parcialmente coerente, introduzindo o fenômeno da interferência e definindo a função de coerência \cite{mandel1995optical,wolf2007introduction,fowles1989introduction}.  A teoria de interferência é baseada no princípio de superposição linear dos campos. De acordo com esse princípio, o campo elétrico em um dado ponto (P) no espaço vazio pode ser entendido como a soma vetorial dos campos que atuam nesse ponto, podendo ser representada como
\begin{equation}\label{eq1.1}
     \boldsymbol{E}(\boldsymbol{r},t) = \boldsymbol{E_{1}}(\boldsymbol{r},t)  + \boldsymbol{E_{2}}(\boldsymbol{r},t).
\end{equation}

\begin{figure}[ht!]
    \centering
    \includegraphics[width=8cm]{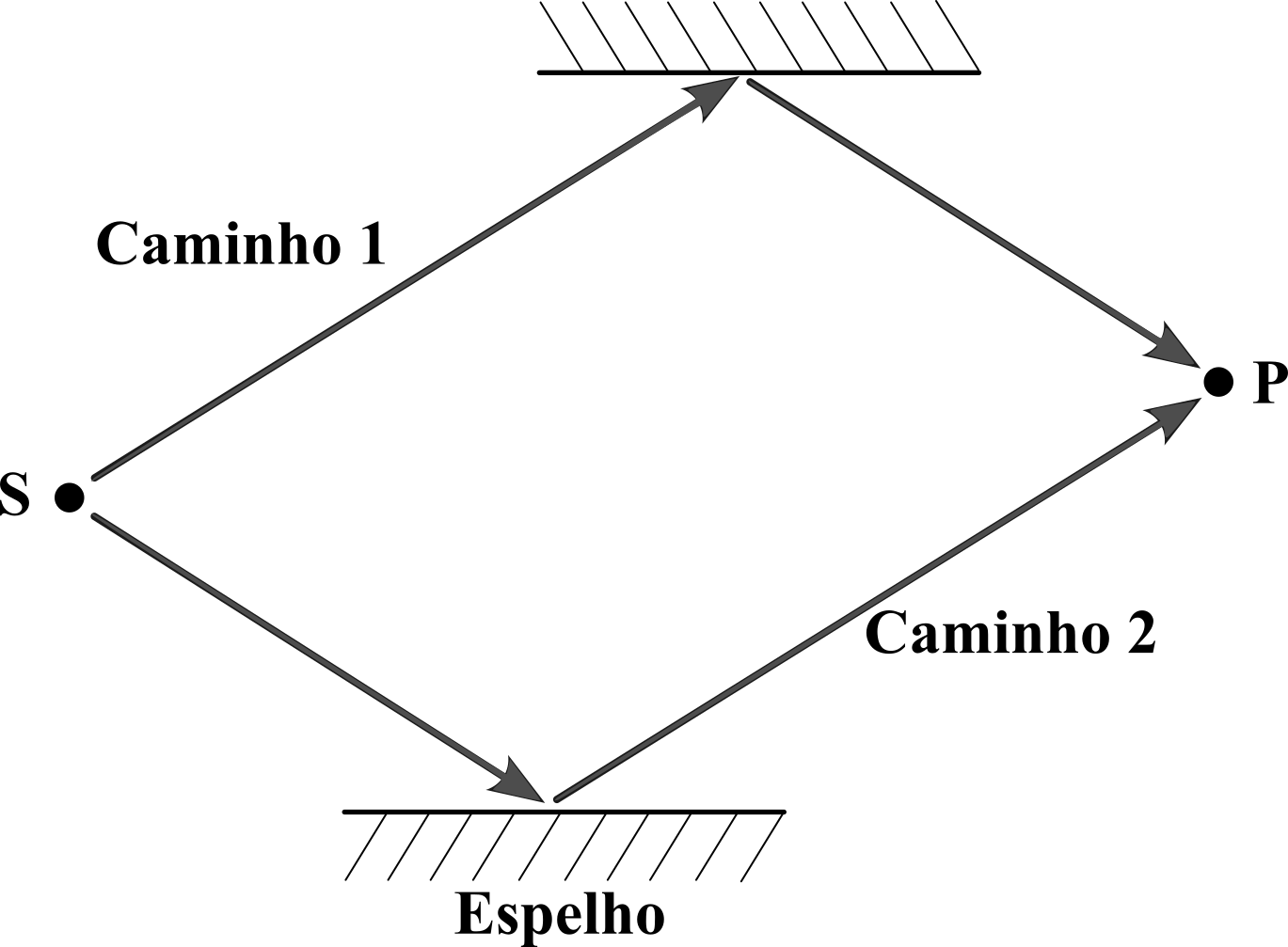}
    \caption{Fonte: Autor}
    \label{fig:00}
\end{figure}

Em geral, no experimento de interferência, os campos $ \boldsymbol{E}_1(\boldsymbol{r},t)$ e  $ \boldsymbol{E}_2 (\boldsymbol{r},t)$ são devidos à mesma fonte (S), a exemplo da Fig. \ref{fig:00}.

Em um dado instante de tempo a intensidade varia rapidamente, portanto, é útil definir a intensidade média do campo expressa por \textcolor{blue}{[Ref]}
\begin{equation} \label{eq1.2}
     I \equiv \langle \boldsymbol{E}^{\dag }(\boldsymbol{r},t) \boldsymbol{E}(\boldsymbol{r},t) \rangle,
\end{equation}
em que $\langle ... \rangle$ representa a média temporal da intensidade e $\dag $ representa o complexo conjugado, considerando que os campos são estacionários\footnote{Uma quantidade estacionária significar que o tempo médio independe da escolha de origem do  tempo \cite{fowles1989introduction}.} e apresentam a mesma polarização.
Ao substituir a equação \eqref{eq1.1} em equação \eqref{eq1.2}, obtemos
\begin{equation}\label{equação3}
\begin{split}
    I & =  \langle \boldsymbol{E_{1}}^{\dag }\boldsymbol{E_{1}} \rangle  + \langle \boldsymbol{E_{2}}^{\dag }\boldsymbol{E_{2}} \rangle +
     2 \Re \langle  \boldsymbol{E}_1^{\dag } \boldsymbol{E}_2 \rangle,\\
    I &  =   I_1 +  I_2  +
     2\Re\Gamma_{12}(\tau) ,
\end{split}
\end{equation}
\noindent onde $ I_1 \equiv \langle|\boldsymbol{E}_1|^2 \rangle $ e  $ I_2 \equiv \langle|\boldsymbol{E}_2|^2\rangle $.
O argumento da parte real, $ \Gamma_{12}(\tau)$ representa a função de coerência mútua ou função de correlação entre os campos $\boldsymbol{E}_1(\boldsymbol{r},t)$ e $\boldsymbol{E}_2(\boldsymbol{r},t)$, eles diferem um do outro devido ao caminho óptico percorrido por cada campo durante sua propagação. Considerando $t$ o tempo que o campo ${E}_1(\boldsymbol{r},t)$ leva para percorrer o caminho 1 na Fig. \ref{fig:00} e $t + \tau $ o tempo que o campo ${E}_2(\boldsymbol{r},t)$ leva para percorrer o caminho 2, $\tau$ é a diferença de tempo devido ao caminho óptico percorrido pelos campos. Podemos escrever a função de correlação explicitamente como
\begin{equation}
    \Gamma_{12}(\tau) = \langle  \boldsymbol{E}_1(\boldsymbol{r},t)^{\dag } \boldsymbol{E}_2(\boldsymbol{r},t+\tau) \rangle.
\end{equation}
De forma mais geral temos
\begin{equation} \label{equação5}
    \Gamma_{ij}(\tau) =\langle  \boldsymbol{E}_i(\boldsymbol{r},t)^{\dag } \boldsymbol{E}_i(\boldsymbol{r}, t+\tau) \rangle, \text{ com i, j = 1, 2}.
\end{equation}
Quando $i=j$ obtemos o produto do campo por ele mesmo e interpretamos esse termo como uma função de autocorrelação.
A função de correlação normalizada, também chamada de grau de coerência, é definida como
\begin{equation}
    \label{eq1.4}
    \gamma_{ij}(\tau)  \equiv \frac{ \Gamma_{ij}(\tau)}{ \sqrt{\Gamma_{ii}(0)\Gamma_{jj}(0) }}  = 
    \frac{ \Gamma_{ij}(\tau)}{ \sqrt{I_{i}I_{j} }}.
\end{equation}

A expressão equação \eqref{eq1.4} obedece $ 0 \leq |\gamma_{ij}| \leq 1 $ sendo
\begin{align*}
       |\gamma_{ij}|  = 1,&  \text{ coerência total}, \\
    0 <|\gamma_{ij}|<   1,&  \text{ coerência parcial}, \\
       |\gamma_{ij}|  = 0,&  \text{ incoerência total.}
\end{align*}

A equação \eqref{equação3} pode ser escrita em termos do grau de coerência, substituindo equação \eqref{eq1.4} em equação \eqref{equação3},
\begin{equation} \label{eq1.5}
    I =  I_1  +  I_2  +
     2\sqrt{I_{1}I_{2}} \Re ( \gamma_{12} ).
\end{equation}
 
Este resultado teórico pode ser verificado experimentalmente por meio da fenda dupla de Young, fazendo uma conexão entre a visibilidade do padrão de interferência e o grau de coerência.
Para isso, definimos a visibilidade das franjas
  \begin{equation} \label{eq1.6}
      V \equiv \frac{  I_{max} - I_{min}  }{  I_{max} + I_{min} },
  \end{equation}
onde $I_{max}$ e $I_{min}$ tratam-se da intensidade máxima e mínima da luz, respectivamente.
Considerando os valores máximo e mínimo da equação \eqref{eq1.5}, respectivamente,
  \begin{equation} \label{eq1.7}
   \begin{split}
      I_{max} &= I_{1} + I_{2} +2\sqrt{I_{1}I_{2}} |\gamma_{12} |, \\       
      I_{min} &= I_{1} + I_{2} -2\sqrt{I_{1}I_{2}} |\gamma_{12}|,
    \end{split}
  \end{equation}
e substituindo equação \eqref{eq1.7} em equação \eqref{eq1.6} conectamos a visibilidade $V$ com as intensidades dos feixes,
\begin{equation}
    V = \frac{ 2\sqrt{I_{1}I_{2}} }{ I_{1} + I_{2} }|\gamma_{12}|.
\end{equation}

Para o caso especifico em que $I_{1}=I_{2}$, obtemos a relação desejada entre a visibilidade e o grau de coerência,
\begin{equation}
    V = |\gamma_{12}|.
\end{equation}

Dessa forma, de maneira simples, podemos verificar o grau de coerência, ou melhor, obter um valor aproximado da função de correlação. Isso ocorre porque, na prática experimental ou na simulação, as fendas não representam exatamente dois pontos, uma vez que possuem uma certa largura.

 \subsection{Densidade espectral cruzada} \label{Densidade espectral cruzada}
No caso dos feixes parcialmente coerentes, não temos uma descrição do campo que possa ser representada por apenas um modo, como um feixe gaussiano, mas sim por um ensemble de modos.
Contudo, podemos descrever o campo por meio das suas correlações, como apresentado na equação  \eqref{equação5}. Por motivos experimentais, é mais vantajoso utilizar a representação desta mesma função no espaço de frequência em problemas envolvendo campos e fontes estacionários aleatórios \cite{wolf2007introduction, mandel1995optical}. Por esse motivo, utilizaremos a Densidade Espectral Cruzada, no inglês, \textit{Cross-Spectral Density} (CSD),
 \begin{equation} \label{eq10}
     W(\boldsymbol{r}_1,\boldsymbol{r}_2,\omega) =  \frac{1}{2\pi}\int^{\infty}_{-\infty}\Gamma(\boldsymbol{r}_1,\boldsymbol{r}_2, \tau) e^{ i\omega \tau }d\tau.
 \end{equation}

A CSD é a transformada de Fourier no espaço de frequência da função de coerência mútua. Ambas devem obedecer à equação de Helmholtz, o que implica serem hermitianas e não negativas, mais detalhes podem ser vistos no capítulo 4 de Mandel e Wolf \cite{mandel1995optical}. A equação \eqref{eq10} pode ser encontrada de uma maneira mais geral por meio do teorema generalizado de Wiener–Khintchine no estudo de um processo estacionário aleatório, tomando um caso específico onde $ \omega = \omega'$, para mais detalhes ver seção 2.5 em \cite{wolf2007introduction}. A CSD normalizada é comumente chamada de \textit{Degree of coherence}(DOC) ou Grau de Coerência Espectral.
\begin{equation}\label{eq11}
     \mu(\boldsymbol{r}_1,\boldsymbol{r}_2) = \frac{ W(\boldsymbol{r}_1,\boldsymbol{r}_2) }{\sqrt{ W(\boldsymbol{r}_1,\boldsymbol{r}_1)W(\boldsymbol{r}_2,\boldsymbol{r}_2) }}.  
 \end{equation}
 
 Note que a dependência com a frequência foi omitida na equação \eqref{eq11} sem perda de generalidade. Em alguns livros e artigos a nomenclatura do grau de coerência é utilizada tanto para a função de coerência normalizada como para a CSD normalizada. É importante deixar claro que as funções de coerência e CSD são ligadas pela transformada de Fourier. No entanto, as funções normalizadas não são necessariamente pares de transformada uma da outra. As condições necessárias para as mesmas serem pares, ou não, pode ser vista com detalhes em \cite{Friberg:95}. Neste trabalho, usaremos o termo ``grau de coerência'' para nos referirmos à função de coerência normalizada e ``DOC'' para nos referirmos à CSD normalizada.

   
 \section{TEORIA: Expansão em modos } \label{TEORIA: Expansão em modos}

Nessa seção iremos apresentar como representar a CSD em termos de expansões e apresentar um exemplo de fonte, o Gaussian Schell-Model(GSM);

\subsection{Expansão em Modos Coerentes}
 
Sabemos que a CSD obedece às propriedades de ser hermitiana e não-negativa,
\begin{equation}
    W(\boldsymbol{r}_1,\boldsymbol{r}_2) = W^*(\boldsymbol{r}_1,\boldsymbol{r}_2)
\end{equation}
e
\begin{equation}
    \sum_{j=1}^{n}  \sum_{k=1}^{n} \  a_j^* a_k W(\boldsymbol{r}_j,\boldsymbol{r}_k)\geq 0.
\end{equation}
 Quando $a_{j,k}$ são constantes reais ou complexas, a CSD se torna um \textit{kernel} do espaço de Hilbert-Schmidt. 
 Isso implica que podemos representá-la como uma expansão em modos ortonormais.
 Algumas vezes essa expansão é interpretada por meio do Teorema de Mercer, mas ambas implicam na mesma expressão, a saber
 \begin{equation} \label{eq1.12}
     W(\boldsymbol{r}_1,\boldsymbol{r}_2)  = \sum_{l=1}^{\infty} \ \lambda_l\ \psi^*_l(\boldsymbol{r}_1)\ \psi_l(\boldsymbol{r}_2).
 \end{equation}
 
 A função ortonormal $\psi(\boldsymbol{r})$ e seu autovalor $\lambda_l$, devem obedecer tanto à equação de Helmholtz como à integral homogênea de Fredholm do segundo tipo,
\begin{equation}
   \int  W(\boldsymbol{r}_1,\boldsymbol{r}_2)\psi_l(\boldsymbol{r}_1) \ d^2\boldsymbol{r}_1 = \lambda_l\psi_l(\boldsymbol{r}_2).
\end{equation}

A equação  \eqref{eq1.12} representa a fonte como uma soma incoerente de modos coerentes, justificando o nome do método: Decomposição em Modos Coerentes(DMC).
Devido a essa característica a DMC é conveniente em alguns casos, pois são utilizados modos coerentes conhecidos como o de Laguerre-Gauss, ou Hermite-Gauss\cite{Starikov:82}.

\subsection{Expansão em pseudos modos} \label{expansao em pseudos modos}
 
 O procedimento para encontrar os autovalores da DMC é demasiadamente complicado, pois se trata de um processo no qual é necessário resolver a equação integral homogênea de Fredholm, o que pode ser bastante difícil dependendo da função escolhida. Essa dificuldade abriu portas para novas ideias, a exemplo das propostas Ref. \cite{Gori:07} e Ref. \cite{Martinez-Herrero:09}. Gori e Santarsiero, em seu trabalho de 2007, mostram uma maneira de construir matematicamente qualquer tipo de função de correlação espacial ou CSD.
 A ideia principal por trás do resultado obtido é levar o estudo de teoria de reprodução de \textit{kernel} no espaço de Hilbert para a teoria de coerência. 
 De modo específico, se temos a CSD que obedece à definição de não-negatividade, podemos escrever, como uma superposição integral da forma
 \begin{equation} \label{eq1.14}
     W(\boldsymbol{r}_1,\boldsymbol{r}_2)\ = \ \int p(\boldsymbol{v}) H^*(\boldsymbol{r}_1,\boldsymbol{v}) H(\boldsymbol{r}_2,\boldsymbol{v}) d^2v,
 \end{equation}
onde $\boldsymbol{v}$ é um vetor de tamanho finito, $H(\boldsymbol{r},\boldsymbol{v})$ é um \textit{kernel} arbitrário e $p(\boldsymbol{v})$ é uma função não-negativa denominada de função peso.
 O próximo passo foi dado em 2009 por Martínez-Herrero et al., que rescreveram a integral da equação  \eqref{eq1.14} como uma expansão em modos
 \begin{equation} \label{eq1.15}
     W(\boldsymbol{r}_1,\boldsymbol{r}_2) = \sum_{l=1}^{\infty} \psi^*_l(\boldsymbol{r}_1) \psi_l(\boldsymbol{r}_2),
 \end{equation}
 onde $\psi_l(\boldsymbol{r})$ são chamados de pseudo-modos. 
Para determinarmos as expressões dos pseudo-modos, temos que resolver 
 \begin{equation} \label{eq1.16}
     \psi_l(\boldsymbol{r}) = \int \sqrt{p(\boldsymbol{v})}\ H(\boldsymbol{r},\boldsymbol{v})\ F_l^*(\boldsymbol{v}) d^2v,
 \end{equation}
  onde $ F_l(\boldsymbol{v})$ são funções ortonormais.  A equação  \eqref{eq1.15} é semelhante ao caso da expansão em modos coerentes, equação \eqref{eq1.12}, mas sem a restrição das funções $ \psi_l$ serem ortonormais. Consequentemente, os pseudo-modos não precisam satisfazer a equação de Fredholm. Dessa forma, o processo fica mais simples do que no caso coerente, pois precisamos saber apenas o \textit{kernel} $H(\boldsymbol{r},\boldsymbol{v})$ e a função peso $p(\boldsymbol{v})$ que geram a CSD. Em seguida, basta substituir na equação  \eqref{eq1.16} e escolher a função adequada (ortonormal) para $ F_l(\boldsymbol{v})$. Esta não é a única maneira de obter os pseudo-modos, há outra maneira baseada no trabalho de Martínez-Herrero, que foi apresentado em \cite{Wang:20}. Nela, a função peso foi reescrita como
 \begin{equation}\label{eq1.17}
    p(\boldsymbol{v}) = \sum^{N,M}_{n,m =0}\ p(\boldsymbol{v}_{n,m})\ \delta(\boldsymbol{v} - \boldsymbol{v}_{n,m}),
 \end{equation}
 $\delta(y)$ trata-se da função delta de Dirac e $\boldsymbol{v}_{n,m} = ( \boldsymbol{v}_{n}, \boldsymbol{v}_{m})$ é uma variável discreta em duas dimensões. Substituindo a equação \eqref{eq1.17} na equação \eqref{eq1.14}, obtemos:
 \begin{equation}\label{eq1.18}
     W(\boldsymbol{r}_1,\boldsymbol{r}_2) = \sum^{N,M}_{n,m} p(\boldsymbol{v}_{n,m})\ H^*(\boldsymbol{r}_1, \boldsymbol{v}_{n,m}) H(\boldsymbol{r}_2,\boldsymbol{v}_{n,m}).
 \end{equation}

 De acordo com Wang, o lado direito da equação \eqref{eq1.18} pode ser lido como uma combinação linear de modos mutualmente incoerentes, na qual $H(\boldsymbol{r},\boldsymbol{v})$ e $p(\boldsymbol{v})$ representam os pseudo-modos e seus respectivos pesos.
 Para o caso em que $N,M \rightarrow \infty$, obtemos a expressão dada na equação \eqref{eq1.14}. 
 A ideia descrita por Wang é basicamente sair do contínuo para o caso discreto, isso implica que estamos considerando um espaçamento maior para a variação de $\boldsymbol{v}$. 
 Por fim, podemos nos questionar qual dos dois caminhos é o melhor para encontrar os pseudo-modos. 
 Os dois caminhos apresentados são equivalentes, mas a maneira apresentado por R. Wang é a mais simples, pois não precisamos resolver a integral na equação \eqref{eq1.16}.
 Alguns artigos apresentam bons resultados utilizando essa metodologia, a exemplo das Refs. \cite{Wang:20,Wang:21,Tian:20,gustavoufsc,Zhu:21}.

\subsection{Exemplos: Fontes GSM}

 O caso mais tradicional e simples, para as fontes 
 parcialmente coerentes,  muito estudado por Emil Wolf \cite{WOLF1978293}
 e F. Gori \cite{GORI1980301} desde 1978, é o modelo GSM, nomeado em homenagem ao trabalho de Allan Carter Schell em sua tese de doutorado \cite{Schell}. Uma forma de obter a CSD para diferentes tipos de feixes é através da equação \eqref{eq1.14}. Apresentada inicialmente por F. Gori, ela se baseia na escolha de um kernel $H(\boldsymbol{r},\boldsymbol{v})$ e de uma função peso 
 $p(\boldsymbol{v})$ apropriada. Possíveis escolhas apresentadas por F. Wang em Ref. \cite{Wang:21} são
 \begin{equation} \label{eq19}
    H(\boldsymbol{r},\boldsymbol{v}) =  \exp{ 
     -\frac{r^2 }{ 4\sigma^2  } + 2\pi i \boldsymbol{r}\cdot \boldsymbol{v}}
 \end{equation}
 e
 \begin{equation} \label{eq20}
    p(v)=  2\pi\delta^2 \exp{ 
     -2 \pi^2 \delta^2 v^2}, 
 \end{equation}
onde $\delta$ e $\sigma$ são constantes reais. Ao substituir as Eqs. \eqref{eq19} e \eqref{eq20} na equação \eqref{eq1.14} encontramos
 \begin{equation}
 \begin{split}\label{eq25.1}
    W(\boldsymbol{r}_1,\boldsymbol{r}_2) &=
    2\pi \delta^2 \exp{
    - \frac{(r_1^2+r_2^2)}{4\sigma} 
     }\times \\
     \int&\exp{ 
     -2 \pi^2 \delta^2 v^2 
    \ +\ 2\pi i (\boldsymbol{r}_2 - \boldsymbol{r}_1 )\cdot \boldsymbol{v}
    }d^2v.
\end{split}
 \end{equation}

O resultado da integral da equação \eqref{eq25.1} é tabelada, expressão (3323-2) em \cite{tabela_de_integral}. Ao resolver a integral e realizar algumas manipulações, obtemos
 \begin{equation} \label{eq1.22}
W(\boldsymbol{r}_1,\boldsymbol{r}_2) = \exp{ - \frac{(r_1^2 + r_2^2)}{4\sigma^2} -\frac{( \boldsymbol{r}_2 - \boldsymbol{r}_1 )^2}{2\delta^2}
     }.
 \end{equation}

A equação \eqref{eq1.22} representa a CSD de uma fonte GSM, e as constantes nos denominadores têm uma interpretação física para o feixe. O $\delta$ é o \textbf{comprimento de coerência}, que representa a distância máxima entre dois pontos para ter correlação de fase, o $\sigma$ é a \textbf{largura do feixe} que delimita a região onde a distribuição de intensidade gaussiana decai até $1/e^2$.


 \subsubsection*{Expansão em modos coerentes}


Após identificarmos a CSD que descreve a fonte GSM, procederemos à sua decomposição em modos coerentes. Embora seja possível encontrar as expressões por meio da resolução da integral homogênea de Fredholm, esse método se revela complexo. Diante disso, alguns autores, como Starikov \cite{Starikov:82}, propõem soluções alternativas manipulando a função que descreve a CSD.

Neste estudo, adotamos a metodologia proposta por Gori \cite{Gori:15}, que realiza a decomposição de modos coerentes para o feixe GSM com twist. A seguir, apresentaremos o mesmo procedimento para o caso do feixe GSM. Para isso, inicialmente, utilizaremos o sistema de coordenadas polares $(r,\theta)$ e reescreveremos a equação \eqref{eq1.22} da seguinte maneira
 \begin{equation}
    W(\boldsymbol{r}_1,\boldsymbol{r}_2)  = \exp{ b r_1 r_2(\tau +1/\tau) - s(r_1^2 + r_2^2 ) },
     \label{eq23}
 \end{equation}
 com
 \begin{equation} \label{eq:definicoes da mudanca de coordenada} 
    \begin{split}
        s &= a + b
        \\
        a &=  1/{4\sigma^2}, 
        \\
        b &= 1/{2\delta^2},
    \end{split}
    \end{equation}
 e
    \begin{equation} \label{eq:tau}
        \tau = \exp{ -i(\theta_1 - \theta_2 )  }.
 \end{equation}
 
 Sabemos que o 1° termo do lado direto da equação \eqref{eq23} pode ser  reescrito, por meio da relação existente entre a função geratriz e  a expansão das funções de Bessel modificada $I_m$ (relação 9.6.33 de \cite{abramowitz1988handbook}), dada por
\begin{equation} 
    \exp{b r_1 r_2 \left( \tau +1/\tau  \right) } = \sum^{\infty}_{m = -\infty}\tau^{m} I_m(2b r_1 r_2).
\end{equation}
  
Substituindo esse resultado na equação \eqref{eq23}, obtemos
\begin{equation} \label{eq25}
    \begin{split}
        W(\boldsymbol{r}_1,\boldsymbol{r}_2) =
        \sum^{\infty}_{m = -\infty} I_m(2b r_1 r_2)\times  \\
        \exp{ - s( r_1^2 + r_2^2 )  - im(\theta_1 - \theta_2 )}.
    \end{split}
\end{equation}

 O resultado obtido na equação \eqref{eq25} não é suficiente, pois queremos construir uma expressão que seja o  produto de modos coerentes, e  em cada modo deve ter apenas uma variável espacial $r_i$, motivando a separação da função de Bessel em dois termos. Todavia, primeiro faremos algumas manipulações para fornecer uma interpretação física mais clara da 
 equação, modificando os componentes da CSD de cada sub-índice $m$ para obter
\begin{equation}\label{eq26}
     W(\boldsymbol{r}_1,\boldsymbol{r}_2) =  \sum^{m = \infty}_{m = -\infty} \mu_m M_m(\boldsymbol{r}_1, \boldsymbol{r}_2),
 \end{equation}
com 

\begin{equation*}
      M_m =
      N_m I_m(2b r_1 r_2)
      \exp{ -s(r_1^2 + r_2^2) - im(\theta_1 - \theta_2)},  
 \end{equation*}
 e
 \begin{equation*}
     \mu_m =  \frac{\pi}{2\sqrt{s^2 -b^2}}\left(\frac{b}{s+\sqrt{s^2 - b^2}} \right)^{|m|}, 
 \end{equation*}
 \begin{equation*}
     N_m =  \frac{2\sqrt{s^2 -b^2}}{\pi}\left(\frac{s+\sqrt{s^2 -b^2}}{b} \right)^{|m|}. 
 \end{equation*}

 A escolha de $N_m$ e $\mu_m$ são feitas de maneira que  $M_m(\boldsymbol{r}_1, \boldsymbol{r}_2) $ seja unitária, para qualquer escolha de $m$. 
 A unitariedade é satisfeita se $ M_m $ obedece à  relação 6.611.4  em \cite{gradshteyn2014table}.  Analisando a expressão de $\mu_m$ é fácil ver que ela deve tender a zero quando ``$m$'' tende a $\pm\infty$, pois o termo do quociente é maior. 
 Logo, a expansão tem um número finitos de termos. Em física, sempre queremos soluções que convirjam, isso implica que devemos ter um número finito na expansão. Além disso, a equação \eqref{eq26} nos permite interpretar a expressão como uma soma incoerente de fontes coerentes, no qual cada fonte é caracterizada por $M_m$ e com peso dado por $\mu_m$.
 
 Ainda não estamos satisfeitos com essa expressão, pois ela não se mostra adequada para uso futuro como um objeto matemático no experimento/simulação. Para que isso aconteça, 
 queremos que $M_m$ seja descrito como produto de um campo com seu complexo conjugado. Com esse objetivo, utilizaremos a relação da função de Bessel modificada de ordem $m$ com o produto de outras duas funções, de modo que as variáveis $\boldsymbol{r}_1$ e $\boldsymbol{r}_2$, que estão agrupadas na função de Bessel, sejam separadas. 
 Podemos obter essa relação utilizando a identidade (8.976.1) de \cite{gradshteyn2014table}.
 
\begin{equation}\label{eq27}
    \begin{split}
        I_{|m|} \left( \frac{ 4\sqrt{\xi} r_1 r_2 }{ (1 - \xi) w^2 }\right) =
        (1-\xi)\xi^{|m|/2}  \left(2\frac{r_1 r_2}{w^2}\right)^{|m|}
        \\ \times 
        \exp{\left( \frac{2\xi}{1-\xi}  \right) 
        \frac{r_1^2 + r_2^2}{w^2}}
         \sum^{\infty}_{n=0}\frac{n!}{(  |m|+n)!}\xi^n
        \\ \times
        L^{|m|}_{n}\left( \frac{2 r_1^2}{w^2} \right) 
        L^{|m|}_{n}\left( \frac{2 r_2^2}{w^2} \right).    
    \end{split}
\end{equation}

com 
\begin{equation}
    \xi = \frac{s  -\sqrt{s^2 -b^2 }}{ s  +\sqrt{s^2 -b^2 }} \label{xixi}
\end{equation}

e

\begin{equation}
    1/w^2  =\sqrt{s^2 -b^2 }.
    \label{1_w}
\end{equation}

Observe que estamos usando a propriedade da função de Bessel modificada, onde $I_{-m} = I_{m} = I_{|m|} $, identidade (9.6.6) em \cite{abramowitz1988handbook}.
Ao realizar algumas manipulações nas constantes da equação \eqref{eq26}, que possibilita a utilização da expressão \eqref{eq27} com \eqref{1_w} e \eqref{xixi} obtemos
\begin{equation}\label{eq28}
     W( \boldsymbol{r}_1, \boldsymbol{r}_2) =
     \sum^{ \infty}_{n = 0} 
     \sum^{m = \infty}_{m = -\infty}
     \lambda_{n.m} 
     \psi_{n,m}^*( \boldsymbol{r}_1,\theta)
     \psi_{n,m}(\boldsymbol{r}_2,\theta^{'}),
 \end{equation}
 onde,
 \begin{equation*}\label{eq29}
    \begin{split}
         \psi_{n,m}( \boldsymbol{r},\theta)=
         \frac{1}{w}
         \sqrt{\frac{2n! }{ \pi(n + |m|)! }}
         \left( \frac{  \boldsymbol{r}\sqrt{2}}{w}\right)^{|m|}
         \\ \times
         L^{|m|}_{n}\left(\frac{ 2\boldsymbol{r}^2}{w^2} \right)
         \exp{- \boldsymbol{r}^2/w^2+ im\theta}
    \end{split}
\end{equation*}
são os modos de Laguerre-Gaus,
\begin{equation}
     \lambda_{n,m}=  \frac{\pi}{s +\sqrt{s^2-b^2}}\left( \frac{s-\sqrt{s^2-b^2} }{s+\sqrt{s^2-b^2}} \right)^{|m|/2+n}, 
     \label{modos de laguerre e gauss}
\end{equation}

A equação \eqref{eq28} representa a CSD da fonte GSM como uma expansão de modos coerentes de Laguerre-Gauss.
Este resultado será útil durante a simulação, pois agora podemos construir a fonte parcialmente coerente utilizando apenas um certo número de modos de Laguerre-Gauss somados incoerentemente. 

\subsubsection*{Expansão em Pseudo-modos}

 Como apresentado na seção \ref{expansao em pseudos modos}, podemos representar a CSD por meio de uma expansão de pseudo-modos. Visando  comparar,  posteriormente, as duas formas de representação para a mesma fonte, necessitamos deduzir a  decomposição da fonte GSM aplicando essa metodologia. Para essa decomposição o processo será mais simples, pois já temos o \textit{kernel} e a função peso, respectivamente, as equação \eqref{eq19} e \eqref{eq20}, substituindo na equação \eqref{eq1.18} e troncando $\boldsymbol{v} \Rightarrow \boldsymbol{v}_{n,m}$ obtemos
 \begin{equation}
     \begin{split}
        W =  \sum^{N,M}_{n,m = 0} 2\pi \delta^2 
        &\exp{   - \frac{  (r_1^2  +  r_2^2) }{4\sigma^2} -2 \pi^2\delta^2 \boldsymbol{v}^2_{n,m} } 
        \\ \times
        & \exp{ 2\pi i ( \boldsymbol{r}_2 - \boldsymbol{r}_1 )\cdot \boldsymbol{v}_{n,m}  }.
     \end{split}
 \end{equation} 
 
 O limite superior e inferior dos valores de $\boldsymbol{v}_{n,m}$ são definidos a 
 partir de $p(\boldsymbol{v}_{n,m})$, e, neste caso em específico, dependem do argumento da exponencial. Em 
 alguns artigos a escolha é dada para quando $\sqrt{p(\boldsymbol{v}_{n,m})} \sim  \exp{- 4}$ 
 \cite{gustavoufsc, Wang:20} e $\sqrt{p(\boldsymbol{v}_{n,m})} \sim  \exp{- 2}$ \cite{Tian:20}. 
 Aqui utilizaremos como parâmetro o valor $ \exp{- 4}$, resultando em
 \begin{equation*}
     \boldsymbol{v}_{n} =  \{ -\sqrt{2}/\pi \delta \ ,\ \sqrt{2}/\pi \delta \ \}.
 \end{equation*}
\section{Simulação: Feixe GSM} \label{Simulacao: Feixe GSM}
Nesta seção iremos apresentar alguns passos que faltam para implementar a simulação, como também uma analise para que a mesma simulação se torne viável dentro dos experimentos. 
\subsubsection*{Experimento para gerar feixes PC}

\begin{figure}[h!]
    \centering
    \includegraphics[width=0.9\columnwidth]{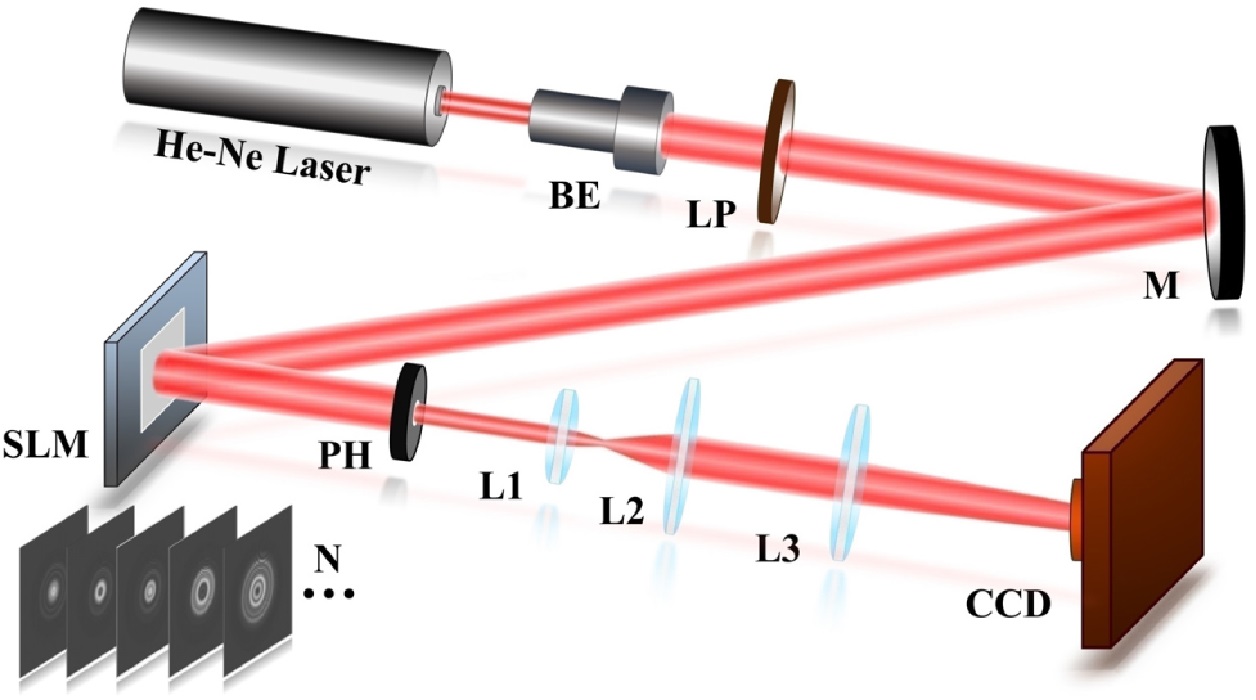}
    \caption{Esboço da configuração experimental. BE, expansor de feixe; LP, polarizador linear; M, espelho refletor;
    SLM, modulador espacial de luz; PH, pinhole; CCD, dispositivo de carga acoplada. Fonte: Wang R. et al \cite{Wang:20}}
    \label{fig:setupesperimental}
\end{figure}

 
Até o momento, falamos sobre a teoria matemática para descrever as classes de feixes parcialmente coerentes. 
Agora, vamos abordar brevemente uma das possíveis abordagens experimentais para geração dos feixes PC, o que nos ajudará a desenvolver a simulação. 
Como exemplos dessa abordagem experimental podemos encontrar na literatura os seguintes trabalhos \cite{Tian:20,Wang:20,WangPengZhangLiuChenWangCai,OLVERASANTAMARIA2019312,F.wang_and_Liu}. 
A ideia central comum a estes trabalhos é bem sintetizada no experimento realizado por Wang, R. et al  \cite{Wang:20}, ilustrado na Fig. \ref{fig:setupesperimental}. Podemos dividir o experimento em três etapas principais:
\begin{itemize}
    \item \textbf{Fonte coerente:} Uma fonte de luz inicial fornece o campo inicial, sendo comumente utilizado um laser devido à sua alta coerência.
    \item \textbf{Manipulação:} Nesta etapa, toda a teoria matemática de decomposição, descrita na seção anterior, é empregada.
    
O feixe proveniente do laser é direcionado para incidir em um modulador espacial de luz (SLM na Fig. \ref{fig:setupesperimental}), cujo papel é modular a fase do campo inicial, para que o feixe de saída (refletido ou transmitido) tenha a propriedade desejada. 
 A modulação, algumas vezes chamadas de máscaras de fases, é devidamente controlada por um computador, elas são escolhidas a partir da decomposição em modos. Cada máscara representa um elemento da série na equação \eqref{eq1.15} e \eqref{eq1.12}. Para obtermos o feixe resultante, as máscaras são inseridas individualmente, modelando o campo e propagando-o até o plano de detecção. Depois que o feixe tem suas fases e intensidades alteradas, a etapa seguinte é rápida devido à velocidade da luz.
 \item \textbf{Detecção:} Por fim, temos o plano de observação, onde um detector/câmera é utilizado para receber as informações do campo transmitido pelo SLM. A câmera deve ficar exposta tempo suficiente para que toda a etapa anterior seja finalizada.
\end{itemize}
     
  As três etapas mencionadas acima oferecem uma visão geral do experimento, que servirá como base para a simulação. No caso do experimento real, há mais elementos ópticos envolvidos entre essas etapas. No entanto, como esses detalhes são mais técnicos e não essenciais para a simulação, optamos por não abordá-los aqui.
  
 \subsubsection*{Simulação de feixes}
  
 Neste artigo, descreveremos o passo-a-passo das três etapas experimentais destacadas na última seção, necessárias para realizar uma simples simulação de um feixe de Laguerre-Gauss. O objetivo é ilustrar ao leitor o processo detalhado para a simulação de um feixe, utilizando como base a biblioteca Light Pipes\footnote{https://opticspy.github.io/lightpipes/}. Além dos \textit{scripts} seguintes, o leitor poderá encontrar versões mais detalhas dos códigos no link \cite{GUBIO_github}.

O primeiro passo é definir as bibliotecas que serão usadas. Isso é realizado nas linha 2 e 3, onde importamos as bibliotecas Light Pipes e Matplotlib. Em seguida, procedemos à definição do tamanho físico de cada pixel, da região de simulação (tamanho da imagem) em  termos do número de \textit{pixels} e do comprimento de onda. Esses parâmetros são essenciais para a criação do campo inicial, que representa uma onda plana. Na linha 10, utilizamos a função \textit{begin} para gerar este campo inicial. 
 
\begin{lstlisting}[language=Python]
########### Bibliotecas ###############
from LightPipes import *
import matplotlib.pyplot as plt
#######################################
# Definindo as propriedade do campo. 
dx = 10*mm      # Tamanho de cada pixel                    
Tamanho = 2**7  # Tamanho da imagem                  
Lambda_ = 620*nm # Comprimento de onda
# Criando o campo
F = Begin(dx,Lambda_,Tamanho)
#######################################
\end{lstlisting}
\begin{figure*}[t!]
    \centering
    \includegraphics[width=1.8\columnwidth]{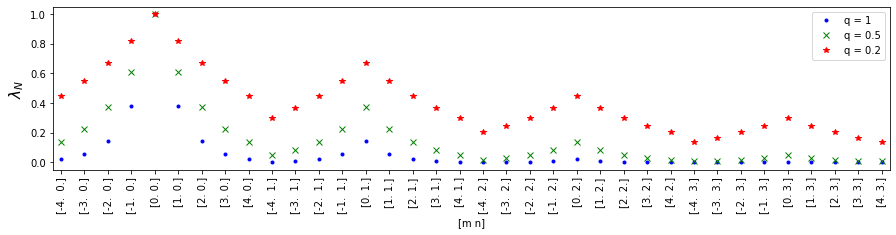}
    \caption{Distribuição de autovalores do feixe GSM para o método de DMC com $n$ e $m$ até 3 e 4, respectivamente.}
    \label{fig:3}
\end{figure*}

Na segunda parte, iniciamos definindo a largura do feixe, conforme indicado na linha 2. Posteriormente, na linha 3, aplicamos uma modulação ao campo inicial usando o modo de Laguerre-Gauss, conforme descrito na Equação \eqref{modos de laguerre e gauss}, com uma ordem radial de n=1 e um valor azimutal |m|=1. Para realizar essa operação, foi utilizada a função \textit{GaussLaguerre}. Em seguida,procedemos à propagação do campo modulado utilizando a função \textit{Forvard} por uma distância de 20 cm. Esta etapa simula a propagação física no experimento até alcançar os objetos ópticos, como por exemplo um interferômetro, uma lente ou uma fenda dupla. 
\begin{lstlisting}[language=Python]
#######################################
Cintura  = 2*mm     # Cintura do feixe  
F = GaussLaguerre(F,Cintura, p=1, l=1,ecs=0) 
F = Forvard(F, 0.2*m)
########################################
\end{lstlisting}

\begin{figure}[h!]
    \centering
    \includegraphics[width=0.5\columnwidth]{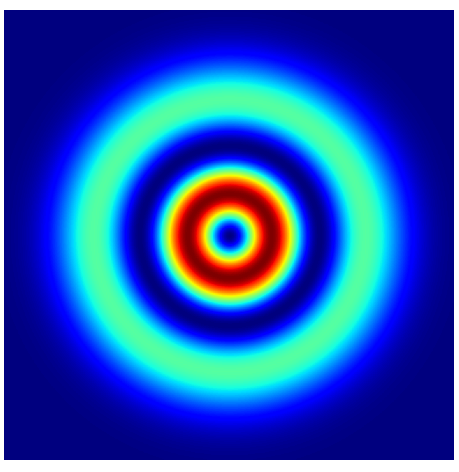}
    \caption{Gráfico da intensidade $I(x,y)$ da simulação de um feixe de Laguerre-Gauss.}
    \label{fig:grafico_da_intensidade}
\end{figure}

Por fim, realizamos uma extração de dados da simulação, semelhante a uma medida física no experimento. Nesse caso, avaliamos a intensidade do campo utilizando a função \textit{Intensity} aplicada ao campo propagante (linha 2). Em seguida, utilizamos a função \textit{plt} para gerar uma representação visual da intensidade do feixe na região da simulação. A figura da intensidade resultante da simulação está retratada na Fig. \ref{fig:grafico_da_intensidade}.
\begin{lstlisting}[language=Python]
#######################################
I = Intensity(0,F)
plt.imshow(I,cmap="jet");plt.axis('off')
plt.show()
#######################################
\end{lstlisting}

\subsection{Modos coerentes}

Para determinar o número de modos necessários na geração dos feixes em cada método, iremos analisar a qualidade do feixe produzido através da distribuição de intensidade e DOC da simulação, comparando com a curva teórica de intensidade de DOC. Na decomposição em modos coerentes (DMC), o número de modos depende estritamente da distribuição da função peso ou autovalor $\lambda_{n,m}$.

Para o feixe GSM, anteriormente obtivemos,
\begin{equation*}
     \lambda_{n,m}= \frac{\pi}{a+b+\sqrt{a^2+2ab}}\left( \frac{a+b-\sqrt{a^2+2ab}}{a+b+\sqrt{a^2+2ab}} \right)^{|m|/2+n}. 
\end{equation*}

Ainda visando simplificar a análise, faremos modificações nas constantes, introduzindo $q = \delta / \sigma$, resultando em $a/b = q^2/2$, como mostram as relações na equação \eqref{eq:definicoes da mudanca de coordenada}. Normalizando a distribuição do autovalor, tomando como referência o primeiro modo, temos
\begin{equation}\label{eq40}
     \lambda_N = \frac{\lambda_{n,m}}{\lambda_{0,0}} =  \left( \frac{1 + q^2/2 - q/2\sqrt{q^2 + 4} }{1 + q^2/2 + q/2\sqrt{q^2 + 4} }\right)^{|m|/2+n}. 
\end{equation}

\begin{figure*}[t!]
    \centering
    \includegraphics[width=1.3\columnwidth]{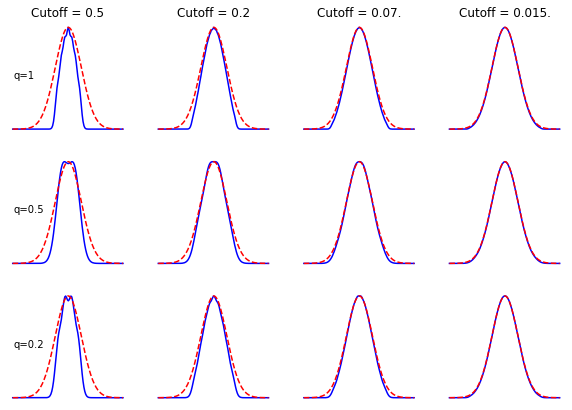}
    \caption{Gráfico da intensidade $I(x,0)$ da simulação (linha azul) e teórico (linha vermelha) para a decomposição em modos coerente. Em cada coluna temos diferentes valores de truncamento e na linhas diferentes comprimento de coerência. }
    \label{fig:grafico_do_cutoff}
\end{figure*}

Uma vez fixado o valor do parâmetro $q$ para um feixe, podemos calcular a distribuição dos autovalores, variando $n$ e $m$. A vantagem de utilizar a equação \eqref{eq40} em termos de uma razão está associada à generalização da análise, pois essa relação deve ser mantida para diferentes valores da cintura do feixe e comprimento de coerência, desde que os parâmetros sejam mantidos na mesma proporção. Isso resulta em uma análise mais abrangente.

Ao analisarmos a figura \ref{fig:3}, observamos que valores de $q$ próximos a 1 tendem a decair mais rapidamente em comparação com valores menores de $q$. Isso indica que necessitamos de menos modos para gerar o feixe com uma região de coerência igual ou próxima à da cintura do feixe, pois as contribuições de ordens maiores vão tornando-se aproximadamente zero mais rapidamente. Isso é uma consequência do fato de o parâmetro $q$ ser diretamente proporcional ao comprimento de coerência; uma CSD com pequeno valor de $\delta$ em comparação à cintura do feixe é também mais incoerente, logo são necessários mais modos incoerentes para reproduzi-la. Apesar de visualizarmos apenas os termos de ordem até $n$ = 3 e $m$ = $\pm$4 na figura \ref{fig:3}, temos contribuições de ordens maiores, com valores cada vez menores.

Ao realizar o cálculo para $q$ = 1, $n$ = 19 e $m$ = 2, obtemos $\lambda_N = 5.72 \times 10^{-25}$, um valor significativamente menor em comparação ao caso com $n = m = 0$ ($\lambda_N = 1$). No caso limite, quando $m$ e $n$ $\sim \infty$, $\lambda_N \sim 0$. O número de modos é finito, pois os autovalores convergem a zero no infinito. Contudo, temos muitos termos de ordens altas que são desprezíveis. Isso nos motiva a realizar um truncamento para eliminar os valores de $\lambda_N$ que são desprezíveis. A escolha do valor do truncamento é baseada em tentativa e erro; devemos testar diferentes valores de truncamento para verificar se o resultado da simulação corresponde ao teórico e, posteriormente, utilizá-lo nos experimentos. Em alguns casos, já foram determinados valores de truncamento a partir de resultados experimentais, como no estudo de Bhattacharjee, que gerou experimentalmente os feixes GSM utilizando a decomposição de Hermite-Gauss, com o valor do truncamento igual a $0.07 \times \lambda_{00}$ \cite{Bhattacharjee}. Faremos uma análise semelhante por meio da simulação, variando o parâmetro $q$ e observando a intensidade para diferentes valores de truncamento (\textit{cutoff}). Antes de realizar essa análise, precisamos aprender a construir a intensidade resultante da soma incoerente, objetivo da próxima seção.


\subsubsection*{Intensidade}


Na seção \ref{TEORIA: Expansão em modos}, descrevemos como encontrar a expansão dos feixes usando dois métodos diferentes. Também apresentamos de forma concisa o modelo do \textit{setup} experimental utilizado na simulação. Nesta segunda etapa, é necessário modelar o campo inicial com a fase desejada, definida pela decomposição em modos. No caso da decomposição em modos coerentes, cada máscara holográfica utilizada no experimento e simulação é definida como
\begin{equation} \label{eq41}
     \Phi_{n,m}(\boldsymbol{r}) = \sqrt{\lambda_{n,m}}\psi_{n,m}(\boldsymbol{r},\theta).
\end{equation}

A expressão da equação \eqref{eq41} indica que, para cada valor de $n$ e $m$, temos uma máscara que modela o campo inicial. Opttamos por usar os modos de Laguerre-Gauss, com seus autovalores definindo um peso. A intensidade de cada termo é dada por
\begin{equation}
    I_{n,m} = | \Phi_{n,m}(\boldsymbol{r}) |^{2} =  \lambda_{n,m}|\psi_{n,m}(\boldsymbol{r},\theta)|^{2}.
\end{equation}

Somando todas as componentes, obtemos o campo final, dado por uma soma incoerente (soma de intensidade) no campo distante, cuja a intensidade teórica \cite{CAIbook} para a fonte GSM é   
\begin{equation}\label{eq42}
    I = \exp{ - \frac{r^2}{4\sigma}}.
\end{equation}

Para ilustrar ao leitor como obter a intensidade do feixe GSM com a decomposição na simulação, apresentamos um exemplo na Ref. \cite{GUBIO_github}. Neste exemplo, definimos um valor máximo da ordem da decomposição e utilizamos todos os termos da série. Também, apresentamos os códigos para a construção da figura \ref{fig:grafico_do_cutoff} com a análise da comparação da distribuição de intensidade simulada com os valores teóricos obtidos através da equação \eqref{eq42}. Na figura \ref{fig:grafico_do_cutoff}, observamos que para um valor de truncamento de $0.015$, é obtida uma boa concordância entre a curva teórica de intensidade e a simulada. 

\subsubsection*{Grau de Coerência}


\begin{figure}[ht!]
    \centering
    \includegraphics[width=0.9\columnwidth]{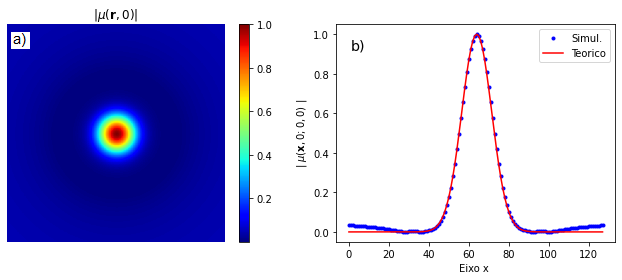}
    \includegraphics[width=0.9\columnwidth]{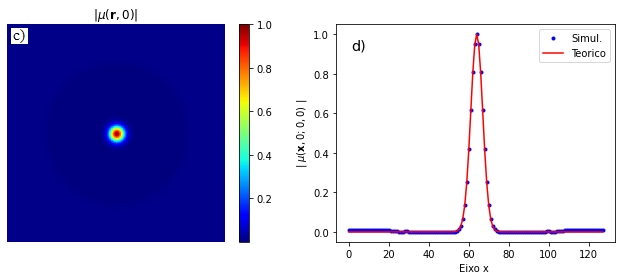}
    \caption{a) e c) são as DOC da simulação com $r_{2}=0$, $q$ = 0.5 e $q$ = 0.2, respectivamente. b) e d) são os DOC da simulação com $r_{2} = y_{1} =0$ e $q$ = 0.5 e $q$ = 0.2, respectivamente, com adição da curva teórica.}
    \label{fig:4}
\end{figure}

Na subseção \ref{Densidade espectral cruzada}, falamos da função DOC que é essencialmente a CSD normalizada. A DOC desempenha um papel crucial na descrição do feixe, pois fornece informações sobre as correlações do campo em dois pontos distintos, $r_{1}$ e $r_{2}$. Dessa forma, é esperado que a simulação se aproxime adequadamente da expressão teórica da DOC para cada feixe. Para realizar essa análise, começaremos determinando a expressão teórica da DOC para o feixe GSM. 
 
Substituindo a equação \eqref{eq1.22} na equação \eqref{eq11}, obtemos
 
\begin{equation}
    \mu(\mathbf{r}_{1},\mathbf{r}_{2}) = \exp{ \frac{(\mathbf{r}_{2}-\mathbf{r}_{1})^2}{ 2\delta^2 } }.
    \label{eq44}
\end{equation}

Na Fig. \ref{fig:4}, notamos que a simulação se aproxima bem da DOC da fonte GSM com a escolha de \textit{cutoff} de $0.015$, tanto para o valor de $q = 0.5$, utilizando 15 modos diferentes, quanto de $q= 0.2$ utilizando 253 modos (código na Ref. \cite{GUBIO_github}).

\begin{figure}[ht!]
    \centering
    \includegraphics[width=1\columnwidth]{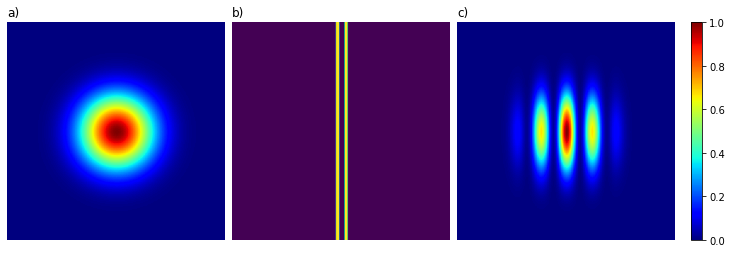}
    \caption{a) Intensidade do feixe GSM para DMC com $z$ = 0 e $q = 0.5$ . b) Fenda Dupla. c) Distribuição de intensidade do padrão de interferência. }
    \label{fig:9}
\end{figure}

\subsubsection*{Fenda dupla}

 Um método experimental comumente empregado para determinar o grau de coerência de um feixe é através da fenda dupla, onde observamos os máximos e mínimos de intensidade no padrão de interferência para calcular a visibilidade. Portanto, na simulação, faremos o feixe GSM passar pela fenda dupla e, em seguida, calcularemos o valor médio da visibilidade utilizando métodos disponíveis em Ref. \cite{GUBIO_github}.
\begin{figure}[ht!]
    \centering
    \includegraphics[width=0.8\columnwidth]{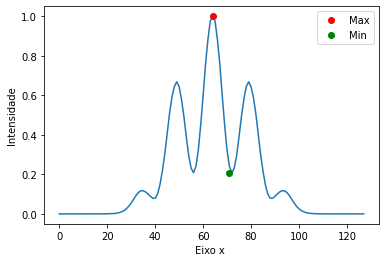}
    \caption{Padrão de interferência com $y$ = 0 de um feixe GSM }
    \label{fig:6}
\end{figure}

Na Fig. \ref{fig:9}-a), podemos visualizar o feixe GSM em $z$ = 0, enquanto na Fig. \ref{fig:9}-b) temos a representação da fenda dupla e, em seguida, o padrão de interferência após a propagação até uma distância $z$, na Fig. \ref{fig:9}-c). Para o cálculo da visibilidade, necessitamos identificar o máximo e o mínimo do padrão de interferência. Diferentemente do padrão de interferência de feixes coerentes, os mínimos de intensidade não são uniformes quando lidamos com feixes incoerentes. Logo, é importante especificar que, no cálculo da visibilidade para feixes parcialmente coerentes, o mínimo considerado deve ser o primeiro em relação ao máximo de intensidade, como exemplificado na figura \ref{fig:6}.  Com $q = 0.5$, obtemos o valor médio da visibilidade igual a $0.63$ em $y =0$ (corte horizontal central da figura  \ref{fig:9}-c ), caracterizando assim um feixe parcialmente incoerente. 


\subsection{GSM em Pseudo-modos}

\begin{figure*}[ht!]
    \includegraphics[width=0.9\columnwidth]{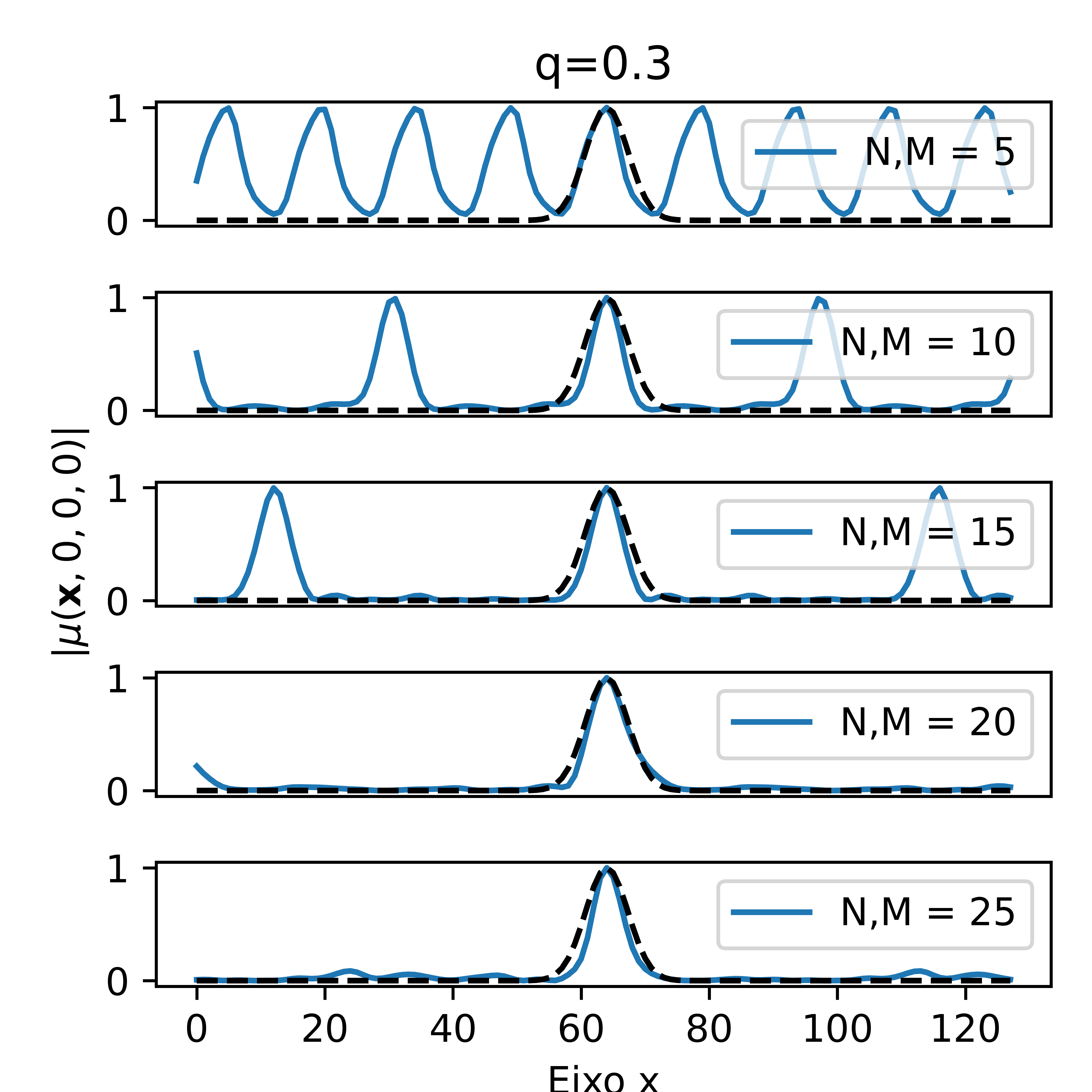}
    \includegraphics[width=0.9\columnwidth]{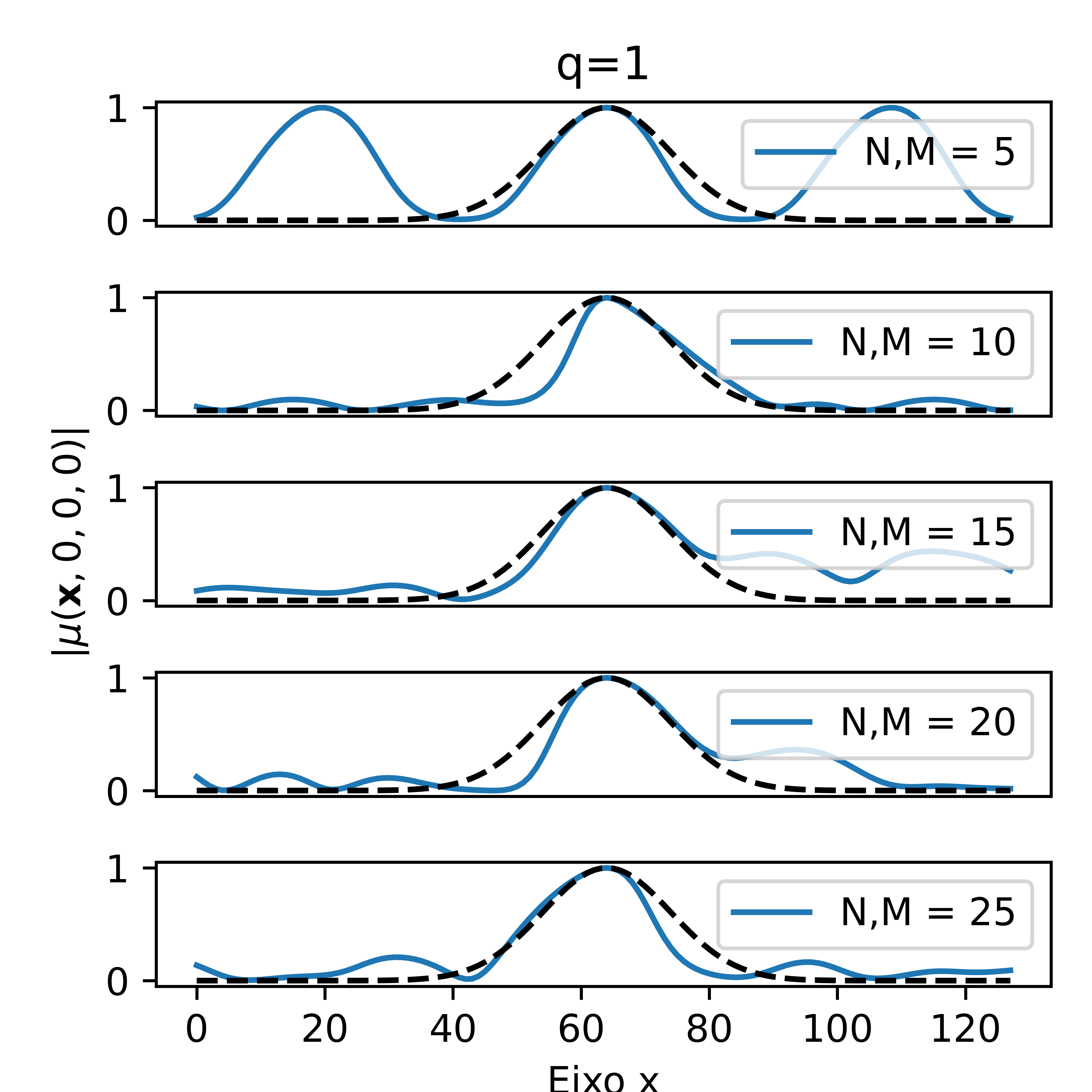}
    \caption{No eixo y temos a função DOC$(x,0;0,0)$ com $l$= 50  variando N e N de 5 a 25 nas curvas em azul. Na linha preta temos a curva teórica descrita pela equação \ref{eq44}. No lado esquerdo temos o caso com $q = 0.3$ e na direita temos $q = 1.0$. }
    \label{fig:8}
\end{figure*}

Para simular o feixe GSM utilizando a decomposição em pseudo-modos, devemos identificar quais são os parâmetros que devem ser ajustados na simulação para obter uma boa aproximação da simulação com a teoria. A metodologia para analisar o número de modos difere do caso coerente, pois não temos funções de pesos bem definidos em termos de autovalores. O procedimento apresentado é inspirado na referência \cite{Tian:20,Wang:20}, no qual se obtém um valor aproximado do número de modos observando a função DOC teórica equação \eqref{eq44}.  Nesse método, temos que adicionar uma fase aleatória no produto entre a função peso e o \textit{kernel} na máscara holográfica, obtendo
\begin{equation} \label{eq45}
    \Phi_{l} = \sum_{n}^{N}\sum_{m}^{M} \sqrt{p(\boldsymbol{v}_{n,m})} H(\boldsymbol{r},\boldsymbol{v}_{n,m})e^{-i\chi_{n,m}^{l}},
\end{equation}
onde $\chi_{n,m}^{l}$ é um número aleatório para cada valor de $l,n$ e $m,$ sorteado no intervalo de [0,$2\pi$]. O subíndice $l$ representa a quantidade de campos somados incoerentemente entre si.  Substituindo as expressões Eqs. \eqref{eq19} e \eqref{eq20} em equação \eqref{eq45}, obtemos

 \begin{equation}
     \begin{split}
        \Phi_{l}(r,v)  =  \sum_{n,m}^{N,M} & \sqrt{2\pi\delta^2} 
       \exp{   -\pi^2 \delta^2 v_{n,m}^2 - r^2/4\sigma^2} 
\\ \times &  \exp{ 2\pi i \boldsymbol{r}\cdot \boldsymbol{v}_{n,m}-i\chi_{n,m}^{l} }.
        \label{eq46}
    \end{split}
\end{equation}

 Nessa abordagem, o método para gerar cada campo da série da equação \eqref{eq1.15} é utilizando a expressão equação \eqref{eq46}, somando coerentemente cada termo de $n$ e $m$ para obter um pseudo-modo $\psi_{l}(r,v)$, em seguida, devemos repetir o processo $l$ vezes.  Como apresentado por Fei Wang, a ideia principal da representação modal é expressar a CSD em termos de uma soma incoerente de modos \cite{Wang:21}. Cada termo da soma incoerente é dito um modo na representação utilizada.  Em contraste com a decomposição coerente, na qual cada elemento da expansão corresponde a um campo distinto, na representação modal todos os termos são idênticos, como interpretado pela expressão equação \eqref{eq45}. Assim, o termo de fase aleatória desempenha um papel crucial na soma incoerente, conferindo distinguibilidade entre os modos.
 \begin{figure}[ht!]
    \centering
    \includegraphics[width=0.9\columnwidth]{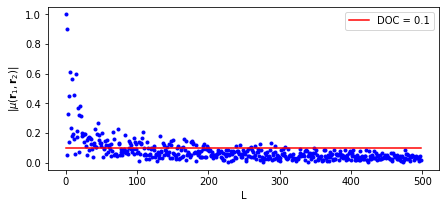}
    \caption{Módulo do função DOC em termo no número de modos incoerentes}
    \label{fig:18}
\end{figure}

 O procedimento para realizar a simulação deve obedecer às etapas do método anterior, alterando apenas a forma de como gerar os modos.  Podemos realizar uma análise qualitativa para identificar quais são os valores adequados de $n$, $m$ e $l$ para a decomposição equação \eqref{eq46}. 
 
 Para determinar os valores de $n$ e $m$, analisamos a função DOC com $r_{2}=(0,0)$ e  $r_{1}=(x,0)$. Na Fig. \ref{fig:8}, apresentamos o resultado dessa simulação para $q=0.3$ e $q=1.0$, variando $N$ e $M$ de $5$ até $25$. Nessa figura, também incluímos a curva teórica da função DOC, conforme definida pela equação \eqref{eq44}. Podemos observar que em algumas situações a função DOC não segue o esperado teoricamente devido ao número baixo de $N$ e $M$. No entanto, ao aumentarmos $N$ e $M$, conseguimos diminuir essas discrepâncias. As oscilações observadas na simulação indicam a presença de várias regiões de coerência no feixe, o que não é desejado para nosso propósito. Para mitigar essas oscilações, é necessário aumentar os valores de $n$ e $m$, introduzindo mais fases aleatórias de modo que, em média, as fases nesses pontos se cancelem. Com base nas curvas apresentadas na Figura \ref{fig:8}, observamos que valores de $N,M \geq 20$ são suficientes para eliminar a grande maioria das regiões de coerência. Além disso, diversos estudos experimentais obtiveram resultados satisfatórios com valores acima de $N$ e $M$ de 15, dependendo das características específicas do \textit{kernel} e do peso utilizados \cite{Tian:20,Wang:20, gustavoufsc}. 
\begin{figure}[ht!]
    \centering
    \includegraphics[width=0.9\columnwidth]{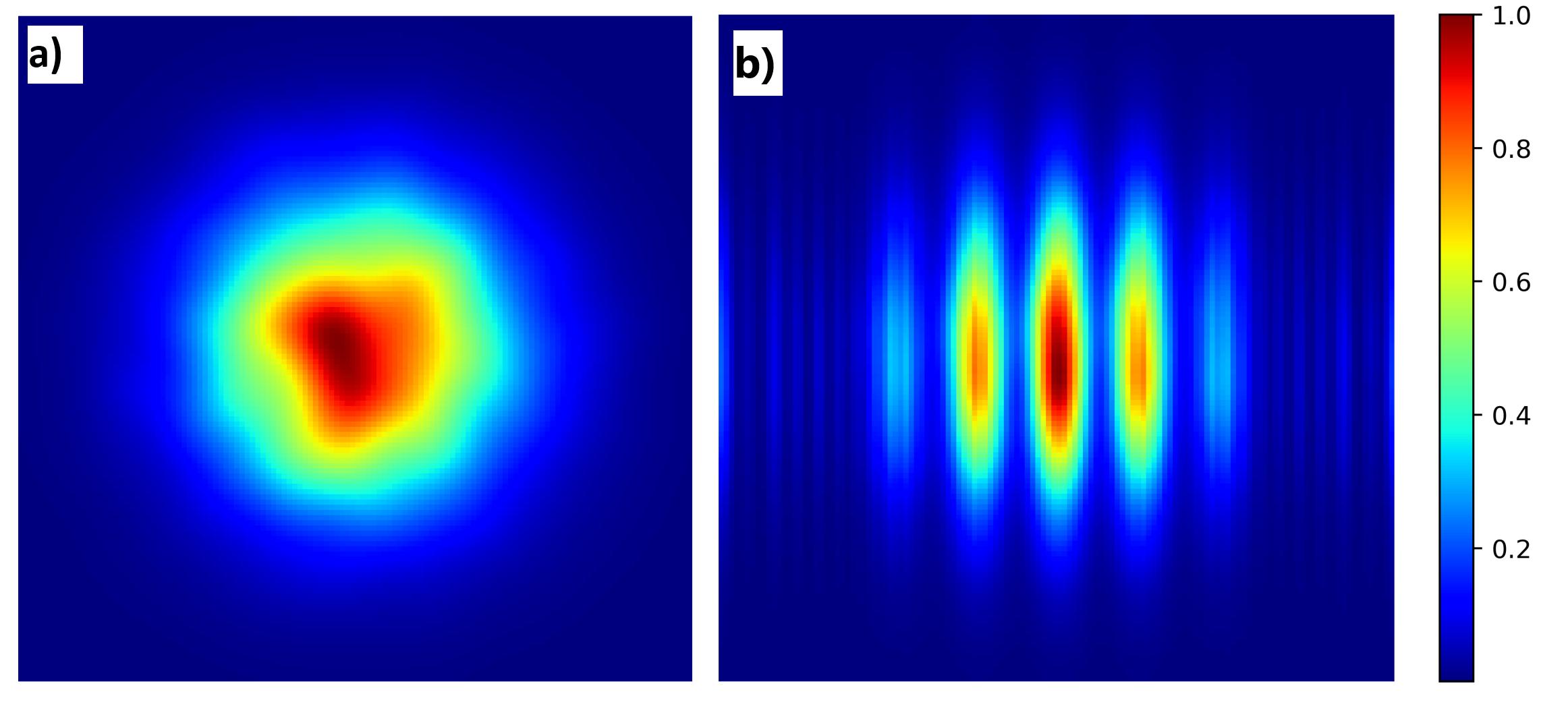}
    \caption{Feixe GSM utilizando pseudo-modos com $q =0.5$. a) Intensidade $z$ = 0.b) Padrão de interferência. }
    \label{fig:22}
\end{figure}
 Para determinar o valor de $l$, analisamos o comportamento do módulo da função DOC em relação ao aumento de $l$. O módulo da função DOC pode ser encontrado em \cite{Tian:20, Wang:20,CAIbook} como sendo 
\begin{equation}
    |\mu(\boldsymbol{r_{1}},\boldsymbol{r_{2}})| =  \sqrt{  \frac{W^*(\boldsymbol{r_{1}},\boldsymbol{r_{2}})W(\boldsymbol{r_{1}},\boldsymbol{r_{2}})}{
    W(\boldsymbol{r_{1}},\boldsymbol{r_{1}})W(\boldsymbol{r_{2}},\boldsymbol{r_{2}})}   }.\label{modudo_doc}
\end{equation}
Utilizando a equação \eqref{eq1.15}, com os pseudo-modos descritos pela equação \eqref{eq46}, e substituindo na equação \eqref{modudo_doc}, obtemos como resultado
\begin{equation} \label{eq49}
    |\mu(\boldsymbol{r}_1,\boldsymbol{r}_2)| = \frac{1}{l} \sum_{l=1}^l\left| \exp{  i(\chi_{n,m}^{l} -\chi_{n',m'}^{l}) } \right|,
\end{equation}
com $ (m, n \neq m', n')$. 

 A  Fig. \ref{fig:18} exibe o gráfico resultante da expressão equação \eqref{eq49}, demonstrando que, à medida que $l$ aumenta, a função DOC tende a diminuir. Esse comportamento sugere um aumento na incoerência entre os termos da decomposição. Quando $l$ se aproxima do infinito, a DOC converge para zero, caracterizando uma soma totalmente incoerente. No contexto dos pseudo-modos, lidamos continuamente com aproximações e truncamentos, selecionando valores que sejam finitos e realizáveis. Assim como nos modos coerentes, almejamos um número mínimo de pseudo-modos para garantir uma geração eficiente dos feixes. Ao analisarmos detalhadamente o gráfico, observamos que após $l$ = 200, a função DOC atinge valores inferiores a $0.1$. Vários estudos sugerem que a faixa de $l$ entre 200 e 300 proporciona resultados experimentais satisfatórios, resultando em valores baixos para o módulo da função DOC, geralmente entre $0.03$ e $0.06$ \cite{gustavoufsc, Tian:20, Wang:20}. Para uma compreensão mais aprofundada e para reproduzir esses resultados, um material de simulação foi preparado e está disponível na referência \cite{GUBIO_github}.
 
 Por fim, repetimos o procedimento da geração do feixe GSM utilizando o método de decomposição em pseudo-modos e o propagamos pelo experimento de fenda dupla, com o objetivo de validar a simulação calculando o valor médio da visibilidade.
 No exemplo da Fig. \ref{fig:22} são apresentados os dados do feixe GSM gerado usando a decomposição em pseudo-modos para $q = 0.5$, onde o comprimento de coerência é metade do valor da cintura do feixe. No painel esquerdo, Fig. \ref{fig:22}-a), a intensidade do feixe é exibida em $z = 0$, enquanto no painel direito, Fig. \ref{fig:22}-b), observamos o padrão de interferência gerado pela simulação do experimento de dupla fenda. Determinamos que a visibilidade média é de $0.65$, o que caracteriza um feixe parcialmente coerente \cite{GUBIO_github}.
\newline

\section{Conclusões}\label{sec8}
Neste artigo, apresentamos uma introdução abrangente ao estudo de feixes parcialmente coerentes, apresentando o arcabouço matemático necessário para simulações em ambientes de propagação livre, bem como em laboratórios ópticos com moduladores espaciais de luz. Abordamos dois métodos de construção e análise de feixes parcialmente coerentes: a decomposição em modos coerentes e a decomposição em pseudo-modos. Também demonstramos de forma detalhada a aplicação da biblioteca \textit{LIGHPIPES} na simulação de feixes. Ambos os métodos de decomposição e simulação foram exemplificados na geração de um feixe GSM. 

Também exploramos técnicas para determinar a qualidade da simulação obtida e do grau de coerência do feixe gerado.
Utilizamos o método experimental de determinação do grau de coerência por meio da fenda dupla, determinando a visibilidade a partir dos máximos e mínimos no padrão de interferência. 

Em suma, este trabalho não apenas aproxima estudantes da teoria óptica de feixes parcialmente coerentes, mas também oferece uma alternativa educacional inovadora para aqueles sem acesso a laboratórios especializados. Através de simulações numéricas, proporcionamos uma experiência de experimentação digital, promovendo a aprendizagem prática e exploratória na área de óptica, bem como uma experiência direta com a programação aplicada a um problema prático de física. Destacamos a flexibilidade das simulações, que podem ser executadas em plataformas gratuitas como o \textit{Google Colab} \footnote{https://colab.research.google.com/ }, tornando-as acessíveis mesmo em dispositivos móveis.

\section*{agradecimentos}
Este trabalho teve o apoio do Conselho Nacional de Desenvolvimento Científico e Tecnológico (CNPq) e da Coordenação de Aperfeiçoamento de Pessoal de Nível Superior (CAPES) - Código 001

\end{document}